\RequirePackage{fix-cm}
\documentclass[smallextended]{svjour3}       
\smartqed  
\usepackage{graphicx}
\usepackage{epsfig}
\usepackage{amssymb}
\usepackage{amsmath}
\usepackage{setspace}
\usepackage{natbib}
\usepackage{multirow}
\begin{document}

\title{Application of Levy Processes in Modelling (Geodetic) Time Series With Mixed Spectra}

\titlerunning{Application of Levy Processes in Modelling (Geodetic) T.S.}        

\author{J.-P.~Montillet     \and X.~He    \and K.~Yu
}


\institute{J.-P.~Montillet \at
              IDYST, University of Lausanne, Lausanne, Switzerland \\
              SEGAL, University Beira Interior, Covhila, Portugal - ORCID ID 0000-0001-7439-7862
              \email{jeanfi \_ montillet@yahoo.fr}           
\and
           X.~He \at
           School of Civil Engineering and Architecture, East China Jiaotong University, Nan Chang, China -ORCID ID 0000-0001-9956-4380
           \and
          K. Yu \at
          School of Environmental Science and Spatial Informatics, China       University of Mining and Technology, Xuzhou, China - ORCID ID 0000-0001-7710-3073
}

\date{Received: xxxx / Accepted: xxxx}

\maketitle

\begin{abstract}
Recently, various models have been developed, including the
fractional Brownian motion (fBm), to analyse the stochastic properties of
geodetic time series, together with the extraction of geophysical signals.
The noise spectrum of these time series is generally modeled as a mixed
spectrum, with a sum of white and coloured noise. Here, we are interested
in modelling the residual time series, after deterministically subtracting geophysical signals from the observations. This residual time series is then assumed to be a sum of three random variables (r.v.), with the last r.v. belonging to the family of Levy processes. This stochastic term models the remaining residual signals and other correlated processes. Via simulations and real time series, we identify three classes of Levy processes: Gaussian, fractional and stable. In the first case, residuals are predominantly constituted of short-memory processes. Fractional Levy process can be an alternative model to the fBm in the presence of long-term correlations and self-similarity property. Stable process is characterized by a large variance, which can be satisfied in the case of heavy-tailed distributions.  The application to geodetic time series imply potential anxiety in the functional model selection where missing geophysical information can generate such residual time series.
%

\end{abstract}


\section{Introduction}\label{section1.0}

Among the geodetic data, time series of daily position of Global Navigation
Satellite System (GNSS) receiver have been of particular interest for the study of geophysical phenomenon at regional and global scales (e.g., study of the crustal deformation due to large Earthquakes, sea-level rise -\citep{Bock,Herring2016,He}. However, these time series contain white noise and long-memory processes (i.e. coloured noise). The scientific community agrees with the existence of a trade-off in estimating both the stochastic and functional models \citep{He}. More precisely, the choice of the stochastic model directly influences the estimation of the geophysical signals included in the functional model (i.e., tectonic rate, seasonal variations, slow-slip events - \citep{Bock,He}). To name a few, it includes the First Order Gauss-Markov (FOGM) model, the white noise with power-law noise \citep{Williams2003,Williams2004}, the Generalized Gauss Markov noise model (GGM), or the Band-pass noise \citep{Langbein2008,Langbein2019}. The optimal selection of the stochastic model in GNSS time series analysis remains a hot topic in the scientific community \citep{Bock,Herring2016,He,He2018}.
\\ \indent  It is widely accepted in the geodesy community \citep{Book2019} that most GNSS time series contain flicker noise which is non-stationary. In addition, recent studies \citep{Langbein2019,He2018} have also advocated the introduction of a random-walk to model small jumps and residual transient signals which is also a non-stationary stochastic processes. Thus, several studies, e.g.,  \cite{MontilletYu}, proposed the use of the fBm, first developed by \cite{Mandelbrodt}, in order to model long-memory processes. \cite{Botai} and \cite{MontilletYu} focused on modelling (residual) geodetic time series using the family of Levy $\alpha$-stable distributions \citep{Nolan2009}. The application of this family of distribution was supported by the ability to model long-memory processes and the existence of impulsive signals/noise bursts in the data sets suggesting deviations from Gaussian distribution \citep{Botai}.
\\ \indent This work discusses several statistical assumptions (i.e. stationary properties, presence of long-term correlations, Gaussianity of the increments) on the underlying processes in the GNSS time series, justifying the application of the fractional Brownian motion (fBm) and the family of Levy $\alpha$-stable distributions introduced in \cite{MontilletYu}. A significant difference between Gaussian and Levy stable distributions is that the latter have heavy tails and their variance is infinite. This means that much larger jumps or flights are possible for Levy stable distributions, which causes their variance to diverge. Many natural processes follow Levy stable distributions. Therefore this work aims at understanding when  the Levy processes can be applied to model geodetic time series.
\vspace{0.5em}
\\ \indent The next section starts with the definition of the residual geodetic time series, the fBm and the relationship with the Fractional Autoregressive Integrated Moving Average (FARIMA) model. From financial analysis, we introduce the family of Levy processes \citep{Panas} and the assumptions in order to relate to other models (i.e. FARIMA, fBm). Section \ref{section3.0} presents the assumptions  on the use of the Levy processes in the model of the residual time series. To do so, we model the residual geodetic time series as a sum of three random variables (r.v.), with the hypothesis that the third one is a Levy process. It involves some justifications compared with established models in the scientific community developed in Section \ref{Section3.1}. In Section \ref{Section3.2}, we develop a $N$ steps method based on the variations of the stochastic and functional models when varying the time series' length. Section \ref{Section3.3} is an application to simulated and real time series.
Finally, Section \ref{Section3.4} discusses the limits of modelling geodetic time series with Levy processes. 
\section{The Stochastic Properties of the Residual Time Series and the Definition of Levy Processes}\label{section2.0}

\subsection{Model of Residual GNSS Time Series}
\label{sec:2.1}
GNSS time series are generally regarded as a sum of geophysical signals (i.e. seasonal signal, tectonic rate) and stochastic processes (e.g., white noise, coloured noise) \citep{Williams2004,Davies}. Modelling the stochastic processes within the geodetic time series is crucial in order to estimate the geophysical signal parameters with reliable uncertainties (\citep{Book2019} { \it Chapter $1$ and $2$}, \citep{He}). 

 Here, the residual time series are defined as the remaining time series after subtracting deterministically modeled tectonic rate and seasonal components (i.e. the functional model), from the GNSS observations. The functional model of those signals is based on the polynomial trigonometric method   \citep{JianXinLi,Williams2003,Tregoning2009}
\begin{equation}\label{firsteq}
s_0(t) = at+b+\sum_{j=1}^N (c_j\cos(d_jt) + e_j\sin(d_jt))
\end{equation}
with $s_0(t)$ the sum of the tectonic rate  (with coefficient $a$ and $b$ in Eq. \eqref{firsteq}) and the seasonal signal (sum of $cos$ and $sin$ functions in Eq. \eqref{firsteq}) at the epoch $t$. Note that $d_j$ is equal to $2\pi j/N$, and $N$ can be equal up to $7$ \citep{He}. If $x(t)$ is the residual time series after subtracting the GNSS time series ($s(t)$) with the functional model ($s_0(t)$) of the geophysical processes (e.g., seasonal signal, tectonic rate),  it is generally formulated the hypothesis that the residual time series is a sum of a residual signal and a noise. Following  \citep{Williams2003,He,Book2019}, the stochastic noise model is described with the variance: 
%
%
%
%
\begin{equation}\label{Secondeq}
E\{\mathbf{n}^{\dag}\mathbf{n}\} = \sigma_{n0}^2\mathbf{I} + \sigma_{n1}^2\mathbf{J}
\end{equation}
where the vector $ \mathbf{n} = [n(t_1), n(t_2), ..., n(t_L)]$ is a multivariate noise with $t_i$ the time at the $i$-th epoch. Note that $n(t_i) = n_0(t_i) + n_1(t_i)$, with $n_0(t_i)$ and $n_1(t_i)$ the white noise and the coloured noise sample respectively at the $i$-th epoch.  ${\dag}$ is the transpose operator,  $\mathbf{I}$ the identity matrix, $\mathbf{J}$ is the variance-covariance matrix of the coloured noise. Finally, $\sigma_{n_0}^2$ and $\sigma_{n_1}^2$ are the variance of the white noise and coloured noise respectively. Therefore, this type of time series belongs to the family of mixed spectra, where the mixed spectrum results from the sum of the spectra corresponding to the two kinds of noise \citep{LiTH}. Note that the length of the time series $L$ is much larger than the number of frequencies $N$ defining $s_0(t)$. 
\\ \indent In the modelling of GNSS time series, a strong assumption is the so-called Gauss-Markov hypothesis (\citep{Book2019}{ \it Chapter $2$}) which states that the noise is Gaussian distributed and wide sense stationary (WSS). Therefore, we  assume the white noise to be zero-mean and Gaussian, whereas the coloured noise with a mean equal to $\mu_C(t)$, slowly varying with time and satisfying the WSS hypothesis \citep{Kasdin,Haykin}. The distribution of the coloured noise is one of the key objective of this study, making various assumptions on the type of processes generating this noise.
 
  Finally, the residual signal is considered to be the remaining geophysical signals (i.e. seasonal component and tectonic rate) not completely estimated due to the mismodelling of the stochastic properties of the time series and other small amplitude (i.e. sub millimeter) short time duration transient signals (i.e. local signals, subsidence, ... ) \citep{Bos,MontilletEtAl2015,Herring2016,He}.
\subsection{Relationship between the Power-law Noise, fBm and FARIMA}
\label{sec:2.2}
The error spectrum of the GNSS time series is best characterised by a stochastic process following a power-law with index $\beta$. A power-law noise model means that the frequency spectrum is not flat but is governed by long-range dependencies. If the probability density function of the noise is Gaussian or has a different density function with a finite value of variance, its fractal properties can be described by the Hurst parameter ($H$). \cite{Montillet2012} has proposed to use the fractional Brownian motion (fBm) model in order to model the statistical properties of the residual time series. The essential features of this process are its self-similar behavior - meaning that magnified and rescaled versions of the process appear statistically identical to the original - together with its nonstationarity, implying a never-ending growth of variance with time \citep{Mandelbrodt}. It is worth mentioning that a damped version of the fBm exists and known as the Mat\'ern process, defined having a sloped spectrum that matches fBm at high frequencies and taking on a constant value in the vicinity of zero frequency  \citep{Lilly2017}.

  Following the definition of the fBm from \cite{Mandelbrodt},  if  $H < 0.5$, the process behaves as a Gaussian variable (anti-persistent); if $H > 0.5$ the process exhibits long-range dependence (persistent); while the case of $H$ equal to $0.5$ corresponds to a pure Brownian motion (white noise). Previous studies \citep{Mandelbrodt,Montillet2012} showed that $H$ is directly connected with $\beta$ by the relation:
\begin{equation}\label{equationBeta}
\beta = 2H-1
\end{equation} 
With this definition, flicker noise corresponds to $\beta$ equal to $1$ or $H$ equal to $1$, random walk to $\beta$ equal to $2$ or $H$ equal to $1.5$, and white noise to $\beta$ equal to $0$ ($H$ equal to $0.5$). Thus, the random walk and the flicker noise are classified as long-term dependency phenomena \citep{Montillet2012}. Based on the Hurst exponent, one can favor similar approaches as in financial analysis to deal with  modeling stochastic processes. 
\\ \indent Long-memory processes are modeled with a specific class of ARIMA models called   FARIMA by allowing  for non-integer differentiating. A comprehensive literature on the application of FARIMA can be found in financial analysis \citep{Granger,Panas}. This model can generate long-memory processes based on the value of the different values of the fractional index $d$ \citep{Granger}.  When $d$ equal to $0$ it is an ARMA process exhibiting short memory; when $-0.5\leq d<0$ the FARIMA process is said to exhibit intermediate memory or anti-persistence. This is very similar to the description of the Hurst parameter in the fBm. There is a relationship between $d$ and $H$ such as  $H=d+0.5$,  well-known in financial time series analysis in the presence of aggregation processes \citep{Panas}. 


%
%
%
\subsection{$\alpha$ Stable Random Variable and the Levy $\alpha$-Stable Distributions}
\label{sec:2.3}
In financial analysis, several models are used, including the fBm and the fractional Levy distribution \cite{Panas,Wooldridge}.  The fractional Levy distribution models the Levy processes and in particular the general family of $\alpha$ stable Levy processes which can be self similar and stationary. Let us recall the definition of a stable random variable.
\vspace{0.5em}
\\$\bold{Definition}$  \textit{\citep{Nolan2009}, chap. 1, definition, 1.6} A random variable $X$ is stable if and only if $X \overset{d}{=} aZ+b$, where $0<\alpha \leq 2$, $ -1 \leq k  \leq 1$, $a \neq 0$, $b \in \mathbb{R}$ and $Z$ is a random variable with characteristic function $\phi(u)=E \{\exp{(iuZ)}\}$  $= \int_{-\infty}^\infty \exp{(iuz)}F(z)$ $dz$.  $F(z)$ is the distribution function of $Z$. $E\{.\}$ is the expectation operator. The characteristic function is:
%
\begin{equation}\label{stableChar}
\phi(u) = \left\{ 
  \begin{array}{l l}
    \hspace{-0.3em} \exp{(-|u|^\alpha [1-ik \tan{ \frac{\pi\alpha}{2}} (sign(u))])}  & \hspace{0.5em}  if \hspace{0.5em} \alpha \neq 1\\
       \exp{(-|u| [1+ik \frac{2}{\pi} sign(u)])}, & \hspace{0.5em} if \alpha = 1\\
  \end{array} \right.
\end{equation}
%
%
%
Where $sign$ is the signum function,  $\alpha$ is the characteristic exponent which measures the thickness of the tails of these distributions (the smaller the values of $\alpha$, the thicker the tails of distribution are), $k \in [-1,1]$ is the symmetry parameter which set the skewness of the distribution. In general, no closed-form expression exists for these distributions, except for the Gaussian ($\alpha$ equal to $2$), Pearson ($\alpha$ equal to $0.5$, $k$ equal to $-1$) and Cauchy ($\alpha$ equal to $1$, $k$ equal to $0$) distributions. Note that the distribution is called a symmetric $\alpha$-stable if $k=0$ \citep{Nolan2009,Sensors,MontilletYu}. Various methods exist to estimate the parameters \citep{Koutrouvelis,Nolan2009}. In the remainder of this paper, we use the  maximum-likelihood method of \cite{Nikias}.
\vspace{0.5em}

%
%
%
%

  Now, if a stochastic process is self-similar, then one can model it with the fBm (see \citep{Tankov}, Definition 7.1). Following \citep{Weron2005}, the most commonly used extension of the fBm to the $\alpha$-stable case is the fractional Levy stable motion (fLsm). This process is defined by the integral representation (see appendices). The fLsm is $H$-self-similar and has stationary increments, with $H$ the Hurst parameter described before. Note that this definition of the Fractional Levy process is different from \cite{Benassi} which is not a self-similar process. In the remainder, we use the fLsm definition from (\citep{Weron2005}, Eq. (6)- recall in the appendices).
%

  Moreover, the  relationship between the fLsm and the fBm is obtained from their definition when $\alpha=2$ (see appendices). If $H=1/\alpha$, we obtain the Levy $\alpha$-stable motion which is an extension of the Brownian motion to the $\alpha$-stable case. The Gaussian case (Brownian motion) is then obtained with $\alpha = 2$ (see \cite{Weron2005} for a comprehensive definition of the fLsm). Further definitions such as the fractional stable noise can be established with the fLsm, but there are out of the scope of this work.
 
%

 Finally, the family of Levy $\alpha$-stable distributions is of a particular interest in this work as the $\alpha$ index is equal to the inverse of the Hurst parameter, therefore in the particular case of the  fLsm.
\cite{Panas} stated that for $1/\alpha <H$, positive increments tend to be followed by positive increments and long-range dependence (persistence); whereas for $0<H<1/\alpha$ positive increments tend to be followed by negative increments (anti-persistence). As a consequence, this family of distributions should be suited when modeling the residual time series with a large amplitude coloured noise  with long-memory processes. With the previous definition of the FARIMA and the relationship to $H$, one can assume that the FARIMA model is then favoured over the ARMA process in the case of large coloured noise within the time series.   If the white noise is predominant (or $H=1/2$), the time series should be fitted with a Gaussian distribution following our assumptions in Section \ref{sec:2.1}, and the ARMA model is favoured over the FARIMA.
\section{Levy Processes Applied to Geodetic Time Series Analysis}\label{section3.0}
This section models the residual GNSS time series as a sum of three r.v. together with the statistical assumptions.  We then develop a $N$-steps method to verify our assumptions on simulated and real time series.
\subsection{Assumptions on the Residual Time Series and the Three Types of Levy Processes} \label{Section3.1}
The residual time series is here modeled  as a sum of three random variables (r.v.). Namely, it is the sum of a white noise, a coloured noise and a third r.v. It is a similar approach used in previous works looking at the presence of a random-walk component in the stochastic model\citep{Langbein2008,Davies,Langbein2019,He2018}. The stochastic properties of the third r.v. should tell us how well is the choice of our initial models (i.e. functional and stochastic). To recall the definition of the Levy processes in Section \ref{sec:2.3}, we postulate that the third r.v. belongs to the Levy processes. We then list the type of Levy processes \citep{Wooldridge,Tankov} depending on the assumptions on the underlying stochastic process:
%
\begin{itemize}
\item[1-] (Levy Gaussian) 
The Levy process is a Gaussian Levy process if the r.v.
follows the properties of a pure Brownian motion also called a Wiener process (identity variance-covariance matrix, zero-mean, stationary process - \citep{Haykin,Wooldridge}).  That is the special case of the fLsm and fBm with $H=1/2$.
The residual time series is assumed to contain mostly short-term correlations modeled with an ARMA process. The residual time series should be modeled with a Gaussian distribution.
%
\item[2-] (Fractional Levy) The residual time series exhibits self-similarity with possibly long-term correlations. The Fractional Levy process is described  by the model of the fLsm for the specific case reduced  to the fBm (see previous section). The long-term correlation process is mostly due to the presence of coloured noise \citep{He}. As explained in \cite{MontilletYu}, the ratio of the amplitude of the coloured over white noise determines which stochastic model of the residual time series should be the most suitable between the FARIMA and ARMA processes. The residual time series should be modeled with a Gaussian distribution following the Gauss-Markov assumption.
\item[3-] (Stable Levy) The Levy process is a Levy $\alpha$-stable motion. That is to generalize important misfit between the selected (stochastic and functional) model ($s_0(t)$) and the observations. If small jumps (or Markov jumps), outliers or other unknown processes are presents, it results in a distribution of the residual time series potentially (severely) skewed, not symmetric, with possibly heavy tails, hence modeling with a Levy $\alpha$-stable distribution. With the relationship between the Levy $\alpha$-stable motion, the fBm and the FARIMA, we assume that the stochastic properties of the residual time series should be described with the FARIMA, especially in the presence of high amplitude coloured noise.
\end{itemize}
The assumption of modelling jumps as Markov jumps in the residual GNSS time series may not be intuitive, because the general model is a Heaviside step function \citep{Herring2016,He}. Those jumps result from equipment changes (i.e. antenna, radom) to the receiver, sudden events (bumps to the antenna), geophysical nature (co-seismic offsets) and variations in the environment of the receiver occasioning multipath (e.g., growing trees, buildings) \citep{Book2019}. In financial time series, the jumps are often resulting from the randomness of the stock prices and modeled as random-walk. In addition, the presence of temporal aggregation processes can affect the persistence in the time series, and sometimes changing suddenly the mean depending on the amplitude of the processes \citep{Holbrook}. That is why in order to assume a  Levy $\alpha$-stable motion as the underlying stochastic model in geodetic time series, we restrict our assumption to small undetectable offsets, modelling them potentially as random-walk. For a complete discussion about this topic, we invite readers to refer to \cite{Gaz} and  \cite{He}.
%
%
\subsection{The N Steps Process}\label{Section3.2}
Let us describe the functional model and the stochastic noise model described in Equation \eqref{firsteq} and \eqref{Secondeq} as a functional interpretation called $\mathcal{F}(\mathbf{\theta}_1)$ and $\mathcal{G}(\mathbf{\theta}_2)$. The functional model described in Equation \eqref{firsteq} is then equal in functional form as $\mathbf{s}_0 = [s_0(t_1), s_0(t_2), ..., s_0(t_L)]$ $= \mathcal{F}(\mathbf{\theta}_1)$, whereas the stochastic noise model described using the variance-covariance matrix in Equation \eqref{Secondeq} is equal to  $\mathcal{G}(\mathbf{\theta}_2)$. We define $\mathbf{\theta}_1 =[a, b, (c_j,d_j)_{j=\{1,N\}}]$ and $\mathbf{\theta}_2 =[a_{wh}, b_{cl}, \beta]$, the vector parameters for the functional  and stochastic noise model respectively. For simplification, we have not included in the functional model  the estimation of possible offsets in the time series (see Appendix $B$ for the model). Also, $a_{wh}$ and $b_{cl}$ are the amplitude of the white and coloured noise respectively. The stochastic noise model is here based on the sum of a white and power-law noise ($PL+WN$). 
\vspace{0.5em}

 Here, our method is based on varying the length of the time series resulting in the variations of the stochastic and functional models, which they allow classifying the type of Levy process. The variations of the length of the time series should take into account that the coloured noise is a non-stationary signal, and thus the properties (i.e. $b_{cl}$, $\beta$) vary non-linearly. However, varying the length of the time series over several years is not realistic taking into account that real time series can record undetectable transient signals, undocumented offsets and other non-deterministic signals unlikely to be modeled precisely \citep{MontilletEtAl2015}. That is why we restrain the variations of the time series length to $1$ year. 
  
   Let us call the geodetic time series $\mathbf{s}$ $=[s(t_1),..., s(t_L)]$ and $\mathbf{s}$ $=[s(t_1),..., s(t_{L+N})]$ at the first and  $N$-th variations respectively. The method can be described as:
\begin{eqnarray}\label{Nsteps}
\hat{\mathbf{s}} &=& \mathcal{F}(\hat{\mathbf{\theta}_1}) +\mathcal{G}(\hat{\mathbf{\theta}_2}) \hspace{0.5em} (estimated \hspace{0.5em} model) \nonumber \\
1^{st} step: \mathbf{s} &=& [s(t_1),..., s(t_L)]\nonumber \\
 \Delta^1 \mathbf{s} &=& \mathbf{s} - [\mathcal{F}(\hat{\mathbf{\theta}_1}) ]_1  \hspace{0.5em} (residual \hspace{0.5em} T.S.-\hspace{0.5em} 1st \hspace{0.5em} step)\nonumber \\
 &\simeq & [\mathcal{G}(\hat{\mathbf{\theta}_2)}]_1 + \mathbf{res}_1\nonumber \\ 
N^{th} step: \mathbf{s} &=& [s(t_1),..., s(t_{L+N})]\nonumber \\
\Delta^N \mathbf{s} &=& \mathbf{s} - [\mathcal{F}(\hat{\mathbf{\theta}_1})]_N \hspace{0.5em} (residual \hspace{0.5em} T.S.-\hspace{0.5em} Nth \hspace{0.5em} step) \nonumber \\
 &\simeq & [\mathcal{G}(\hat{\mathbf{\theta}_2)}]_N + \mathbf{res}_N
\end{eqnarray} 
where $\hat{.}$ corresponds to the estimated vector or observations. $[. ]_j$ means the $j$-th iteration of the estimated quantity. $\Delta^1 \mathbf{s}$ and $\Delta^N \mathbf{s}$ are the residual time series after the first and $N$-th variation of the length of the time series. $\mathbf{res}_1$ and  $\mathbf{res}_N$ are the unmodeled signals and stochastic processes after the first and $N$-th step respectively.
\\ \indent To recall the assumptions in Section \ref{Section3.1}, the residual time series $\Delta^N \mathbf{s}$ is modeled as a sum of three r.v. corresponding to the white noise, coloured noise and a Levy process. Using $N$ iterations and the definition of the various Levy processes in the previous section (i.e., Levy Gaussian, Fractional Levy and Stable Levy) in the previous section, we make several assumptions on the estimated parameters and selected stochastic models  in order to characterize this third r.v.  Table \ref{Table1} summarises the assumptions for these three cases. We use specific mathematical symbols to differentiate between them.  $\triangleq$ means the equality in terms of distribution. $\simeq$, $\sim$ and $\neq$ are related to the variations  of the estimated parameters of the stochastic model associated with the first and the $N$-th iteration. This variation is calculated using the sum of the difference in absolute value between the parameters between the first and the $N$-th iteration. Then, a percentage is deduced based on the initial value of the parameters (at first iteration).  Now specifically, the symbol $\simeq$ means that there are little differences (less than $3\%$) between the estimated parameters of the stochastic model associated with the first and the $N$-th iteration. The symbol $\sim$ means that we allow bigger differences up to $20\%$ . With much larger values, we use the symbol $\neq$.

  Moreover, the estimation of the model parameters is carried out using the Hector software \citep{Bos}. We have restrained  our processing to the stochastic model corresponding to the flicker noise (with white noise - $FN+WN$) and power-law  (with white noise $PL+WN$). The optimal choice of the stochastic model is a current research topic in GNSS time series analysis including  recent studies such as \cite{He}, \cite{He2018} and \cite{Book2019}. To simplify our study, we have preliminarily applied the method based on the Akaike information criterion developed in  \cite{He2018} on the real time series to select the stochastic noise model. Therefore we have selected real time series with stochastic models $FN+WN$ and $PL+WN$. We are not going to develop further this selection process in this study, but readers can refer to \cite{He2018}.

 Furthermore, the fitting of the ARMA($p$,$q$) and FARIMA($p$,$d$,$q$) model to the residual time series is carried out by maximum likelihood following \cite{Sowell}, varying the lags $p$ and $q$ within the interval $[0,5]$. Note that the fractional parameter $d$ is an output of the software Hector \citep{Bos} when fitting the stochastic model during the $N$ iterations. Also, the ARMA/FARIMA model which best fits the residual time series, is selected in order to minimize the Bayesian Information Criterion (BIC) following \cite{MontilletYu}. Finally, one can wonder if the anxiety in the model selection (ARMA, FARIMA) in presence of heavy-tails can modify the performance of the BIC. This topic is currently debated in the statistical community (see \cite{Panahi}). Large tails should be detected in the fitting of the Levy $\alpha$-stable distribution via the  maximum-likelihood method of \cite{Nikias}. Due to the direct relationship between the index $\alpha$ and $H$, we assume that the FARIMA should be chosen defacto over the ARMA model. 
\begin{table}[!htbp]
 \centering 
\caption{Assumptions on the functional model and the stochastic parameters estimated via $N$ iterations (see,$N$-Step method) to characterize the type of Levy processes within the geodetic time series. The symbols and notations are explained in Section \ref{Section3.2} }
\label{Table1}
\begin{tabular}{|l|l|l|l|}
\hline
{\it Type of Process} & Levy Gaussian & Fractional Levy & Stable Levy \\
 \hline
 {\it Mathematical } &  $[\mathcal{G}(\hat{\mathbf{\theta}_2)}]_1 \simeq [\mathcal{G}(\hat{\mathbf{\theta}_2)}]_N $ & $[\mathcal{G}(\hat{\mathbf{\theta}_2)}]_1 \sim [\mathcal{G}(\hat{\mathbf{\theta}_2)}]_N$ & $[\mathcal{G}(\hat{\mathbf{\theta}_2)}]_1 \neq [\mathcal{G}(\hat{\mathbf{\theta}_2)}]_N$\\
 
 {\it Assumptions }& $[\mathcal{F}(\hat{\mathbf{\theta}_1}) ]_1 \simeq [\mathcal{F}(\hat{\mathbf{\theta}_1}) ]_N$ & $[\mathcal{F}(\hat{\mathbf{\theta}_1}) ]_1 \sim [\mathcal{F}(\hat{\mathbf{\theta}_1}) ]_N$ & $[\mathcal{F}(\hat{\mathbf{\theta}_1}) ]_1 \neq [\mathcal{F}(\hat{\mathbf{\theta}_1}) ]_N$ \\
  \hline
 {\it (Distribution)} $\Delta^1 \mathbf{s} \triangleq$ & Gaussian &  Gaussian & Levy $\alpha$-stable \\
  \hline
{\it Model To Characterize } & ARMA(p,q) & ARMA(p,q) or   &  FARIMA(p,d,q)   \\ 
{\it Processes} & & FARIMA(p,d,q) &\\ 
  \hline
\end{tabular}
\end{table}

\subsection{Application to Simulated and Real Time Series}\label{Section3.3}
\subsubsection{Simulated Time Series}
The definition of the Levy processes together with the assumptions in Table \ref{Table1} are applied to the residual of simulated geodetic time series. The simulations of the geodetic time series follow \cite{Williams2004} and the routines associated with Hector  \citep{Bos}. The estimations of the ARMA and FARIMA models follow Section \ref{Section3.2}.

 We simulate $10$ years long time series fixing $a_{wh}$ to $1.6$ mm, $a$ varying between $[1-3]$ mm/yr, $b$ equal $0$, and $(c_1,d_1)$ equal to $(0.4, 0.2)$ mm/yr. According to Table \ref{Table1}, we vary the amplitude of coloured noise $b_{cl}$ following three scenarios:
 ($A$) from low value (i.e. $b_{cl}<0.1$ mm/$yr^{\beta/4}$); ($B$) intermediate  (i.e. $1 mm/yr^{\beta/4}>$ $b_{cl}>0.1$ mm/$yr^{\beta/4}$); and ($C$) high value (i.e. $1 mm/yr^{\beta/4}<$ $b_{cl}$ $<4 mm/yr^{\beta/4}$). In the case of the large amplitude of the coloured noise, the process is unlikely zero-mean stationary. Also, $\beta$ is equal to $1$ (flicker noise) or $1.5$ (power-law noise) in the simulations.

\begin{figure}

 \caption{Percentage of variations of the estimated parameters included in the stochastic and functional models when varying the length of the time series. $(A)$, $(B)$ and $(C)$ refer to the various scenarios with different coloured noise amplitude.\label{Figure1a}}
 \hspace{-8em}
\includegraphics[width=1.7\textwidth]{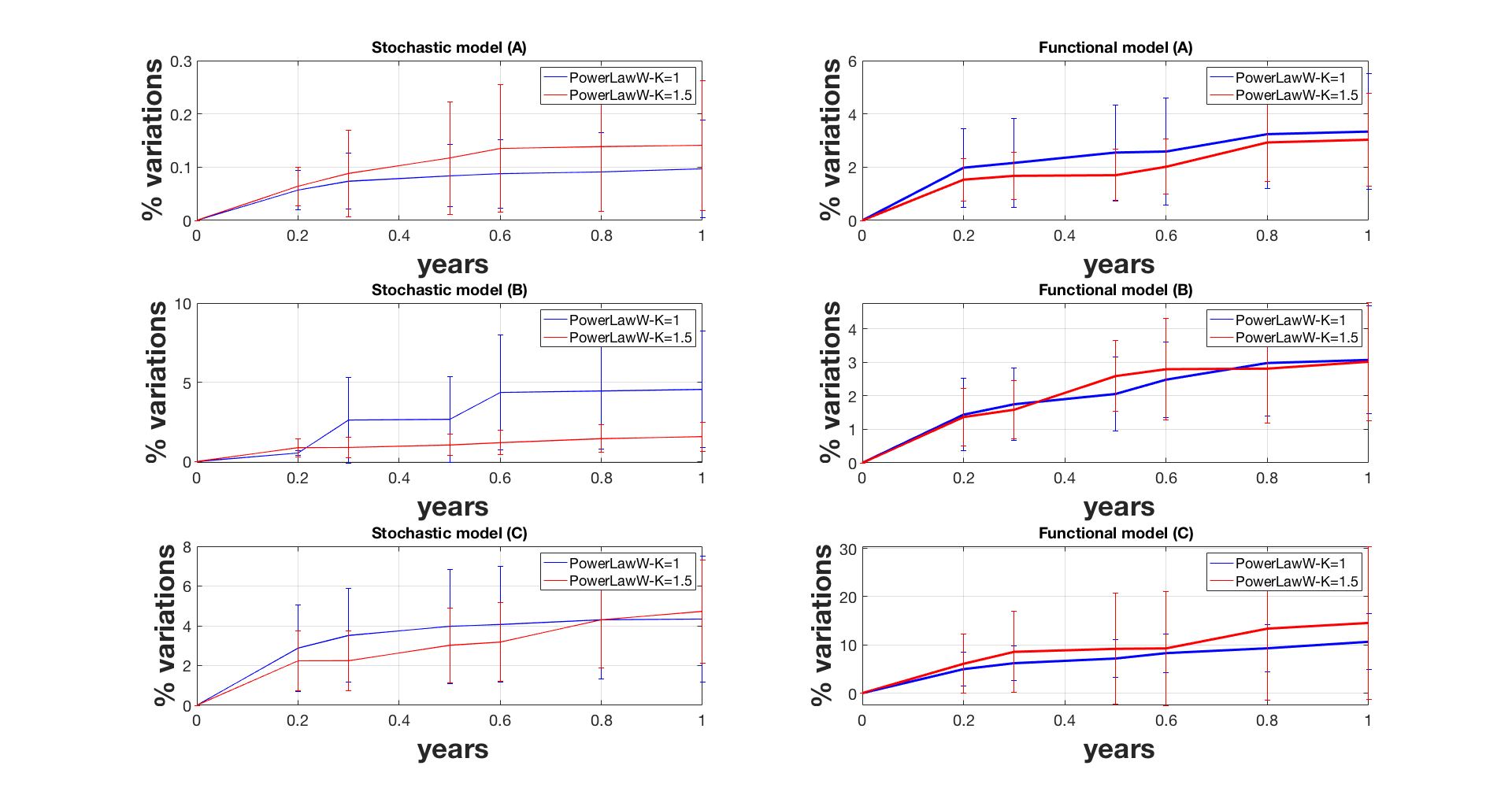}
 
\end{figure} 
 
 Figure \ref{Figure1a}a, \ref{Figure1a}b and \ref{Figure1a}c display the results when averaging over $50$ time series. The variations are in steps of $[0, 0.3, 0.5, 0.7, 0.8, 1]$ year (see X-axis). Each figure corresponds to the different coloured noise amplitude following the three scenarios described above. We show both the variations of the stochastic and functional models. On the Y-axis, these variations are basically the statistics (mean and standard deviation) over the percentage estimated between the parameters of either the stochastic or functional model between the first and $N$-th iteration. For each run, this percentage is calculated using the sum of the differences in absolute value of the various parameters described in Section \ref{Section3.2}. 
 
  In Hector, we use the $PL+WN$ model  \citep{Bos}. The first result which is common to all three figures, is that the variations in the functional model starts earlier than for the stochastic model. Previous studies have shown that there is some part of the noise amplitude absorbed in the functional model \citep{Williams2003,MontilletEtAl2015}. In our scenario, the estimation of the linear trend may fit partially into the power-law noise, hence reducing the variations of the stochastic model. This effect can be amplified with higher spectral indexes. Now, Figure \ref{Figure1a} shows that over $1$ year the variations of the stochastic and functional model are less than $4\%$ (mean value) for small amplitude coloured noise, whereas when increasing the coloured noise the variations increase quickly (e.g., more than $20\%$ for the large coloured noise amplitude for the functional model $(c)$) . Knowing that Hector assumes only stationary signals \citep{Bos}, it means that part of the large variations of the coloured noise is wrongly included in the estimation of the functional model. 

\begin{table}[!htbp]

\caption{Statistics on the Error when fitting the ARMA and FARIMA model to the residual time series following the three scenarios  }
\label{Table2a}
 \centering 
\tiny
\begin{tabular}{|l|l|l|l|l|}
\hline
 {\it Error (mm)} &   & case $A$ & case $B$ & case  $C$  \\
& $\beta$&  $b_{cl}<0.1$ mm/$yr^{\beta/4}$ & $1 mm/yr^{\beta/4}>$ $b_{cl}>0.1$ mm/$yr^{\beta/4}$ & $1 mm/yr^{\beta/4}<$ $b_{cl}$ $<3 mm/yr^{\beta/4}$ \\
 \hline
 {\it ARMA} & $1.1$ & 1.44 $\pm$ 0.01 & 1.74 $\pm$ 0.01 & 1.89 $\pm$ 0.04 \\
  & $1.5$ & 1.46 $\pm$ 0.01 & 1.76 $\pm$ 0.04 & 1.95 $\pm$ 0.05 \\
  \hline
 {\it FARIMA} & $1.1$ & 1.91 $\pm$ 0.02 &1.85 $\pm$ 0.02  & 1.46 $\pm$ 0.02 \\
  & $1.5$ & 1.89 $\pm$ 0.01 & 1.75 $\pm$ 0.03 & 1.59 $\pm$ 0.05 \\
 \hline
 \end{tabular}
\end{table}
\begin{table}[!htbp]
\caption{ Correlation between the distribution of the residuals and the Normal ($Corr.$ $Normal$) and  the Levy $\alpha$-stable distribution  ($Corr.$ $Levy$)  following the three scenarios }
\label{Table2b}
 \centering 
\tiny
\begin{tabular}{|l|l|l|l|l|}
  \hline
  \hline
 {\it Corr. } $[0 -1]$ &   & case $A$ & case $B$ & case  $C$  \\
& $\beta$&  $b_{cl}<0.1$ mm/$yr^{\beta/4}$ & $1 mm/yr^{\beta/4}>$ $b_{cl}>0.1$ mm/$yr^{\beta/4}$ & $1 mm/yr^{\beta/4}<$ $b_{cl}$ $<3 mm/yr^{\beta/4}$ \\

 {\it Corr. Normal }  & $1.1$ & 0.93 $\pm$ 0.14 & 0.92 $\pm$ 0.21 & 0.89 $\pm$ 0.50 \\
 & $1.5$ & 0.92 $\pm$ 0.14 & 0.91 $\pm$ 0.22  & 0.85 $\pm$ 0.31 \\
 {\it Corr. Levy }  & $1.1$ & 0.92 $\pm$ 0.11 & 0.94 $\pm$ 0.14   & 0.96 $\pm$ 0.18\\
 & $1.5$ & 0.93 $\pm$ 0.13 & 0.94 $\pm$ 0.16 & 0.95 $\pm$ 0.18 \\
 \hline
\end{tabular}
\end{table}
 
  Now, Table \ref{Table2a} shows the standard deviation of the difference ($Mean$ $Square$ $Error$) between the ARMA /FARIMA model and the residuals (i.e. $\mathbf{res}_i$ in Equation \eqref{Nsteps}). We do not display any mean, because the fit of the models are done on the zero-mean residuals. Note that the value is averaged over the $50$ simulations, together with the variations of the length of the time series following the same processing as before. The table also displays the averaged correlation between the distribution of the residuals and the Normal or Levy $\alpha$-stable distribution. In agreement with the theory, we can see that the ARMA model fits well residuals with small amplitude coloured noise, whereas with the increase of $b_{cl}$ the FARIMA model fits better than the ARMA model. Looking at Table  \ref{Table2b} in terms of correlation, the Levy $\alpha$-stable  distribution fits  as good as the Normal distribution as long as the distribution of the residuals is Gaussian without large tails or asymmetry. In Section \ref{sec:2.3}, we emphasized that the family of Levy $\alpha$-stable  distributions includes the Normal distribution with specific values of its driving parameters (see Equation \ref{stableChar}). Thus, the results show that for the amplitude of coloured noise, not very large (i.e. Intermediate - case $B$ - in Table \ref{Table2a} and \ref{Table2b}) compared with the white noise, the two distributions show similar results. However, the scenario with large coloured noise amplitude ($C$), which can generate some aggregation processes  thus introducing an asymmetry or large tails in the distribution of the residuals, emphasizes that the family of Levy $\alpha$-stable  distributions perform the best in modelling the residuals' distribution. Note that the asymmetry in the residuals' distribution is relatively limited. Much Larger coloured noise amplitude could produce greater asymmetry in the distribution as seen in financial time series with aggregation processes of high amplitude \citep{Wooldridge}. Finally, those three scenarios support ideally the theory where in the case of small amplitude coloured noise, the stochastic noise properties are dominated by the Gaussian noise, hence supporting a third r.v. defined as a Gaussian Levy. However, the increase of the coloured noise amplitude shows that it is much more difficult to discriminate between the fractional Levy and the stable Levy. The results point out that the third r.v. can be modeled as a stable Levy process when mostly the FARIMA fits the residuals due to large amplitude long-memory processes, hence creating a heavy-tail distribution. This result is restrictive for the application to geodetic time series.
\subsubsection{Real Time Series}
We process the daily position time series of three GNSS stations namely $DRAO$, $ASCO$ and $ALBH$ retrieved from the UNAVCO website \citep{UNAVCO}. The functional model includes the tectonic rate, the first and second harmonic of the seasonal signal, and the occurrence time of the offsets. This occurrence time is obtained from the log file of each station. However, $ALBH$ is known to record slow-slip events from the Cascadia subduction zone \citep{Melbourne}. Thus, we include the offsets provided by the Pacific Northwest Geodetic Array \citep{Miller}. In this scenario we do not know which stochastic model could fit the best the observations. Thus, we use two models: the $PL+WN$ together with the $FN+WN$.

\begin{figure}

 \caption{Percentage of variations of the estimated parameters included in the stochastic and functional models when varying the length of the daily position GNSS time series corresponding to the stations $DRAO$, $ASCO$ and $ALBH$. The statistics are estimated over the East and North Coordinates \label{Figure1aX}}
\hspace{-8em}
\includegraphics[width=1.7\textwidth]{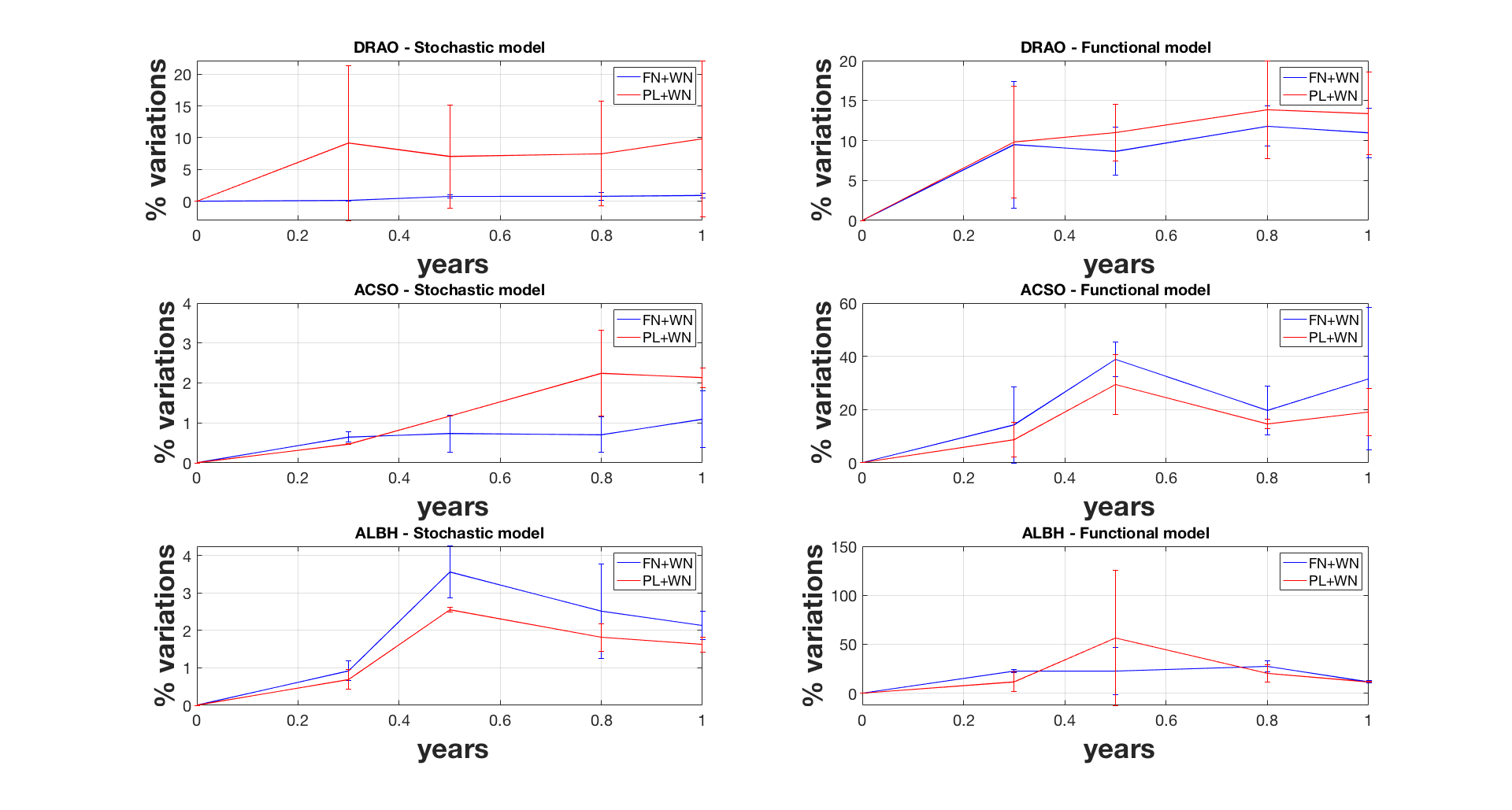}
\end{figure}

 Similar to the previous section, Figure \ref{Figure1aX} displays the percentage of variations of the stochastic and functional models averaged over the East and North coordinates of each station. Note that the average over the three coordinates is displayed in the appendices (see Figure \ref{Figure1XX}). Because the Up coordinate contains much more noise than the East and North coordinates \citep{Williams2004,Montillet2012}, it amplifies the variation of both stochastic and functional models to several order of magnitude, hence overshadowing the results over the East and North coordinates.
 
 Looking at Figure \ref{Figure1aX}, the first result is that for all the stations, there is a strong dependence with the selected noise model. When selecting the power-law noise over the flicker noise model, there is an additional variable to estimate (i.e. the power-law noise exponent $\beta$ in Equation \eqref{equationBeta} ) within the  stochastic noise model. Even though our results show a relationship between modelling the residuals and the choice of the stochastic model, our current work does not deal with this issue. Readers interested in this topic can refer to \cite{He,He2018}.

 The second result is the large variations of the functional model compared with the stochastic model. As explained in the simulations, the functional model partially absorbs the variations of the noise, i.e. the tectonic rate partially fits into the power-law noise. In addition, to some extend at $ASCO$, the sudden increase in the functional model variations at $0.5$ year may be explained due to the absorption of some of the noise with the second harmonic of the seasonal signal. 

 When comparing the variations of the stochastic and functional models with amplitude below $20\%$ for the stations $DRAO$ and $ASCO$, the results agree with the definition of the fractional Levy process defined in Table \ref{Table1} as third r.v. modelling the residuals of the East and North components. The variations of the functional model associated with $ALBH$  are much larger than the other two stations, especially for the $PL+WN$ model with variations up to $50\%$. Those large variations can be explained due to the slow slip events and the difficulty to model the post-seismic relaxations between two consecutive events. In  \cite{He2018}, the authors justified the selection in the stochastic noise model of a random-walk component together with a $FN+WN$ in order to model the mismatch between the functional model and the observations.
 
 \begin{table}[!htbp]
 \centering 
\caption{Statistics on the Error when fitting the ARMA and FARIMA model to the residual time series for each coordinate of the stations $ALBH$, $DRAO$ and $ASCO$ based on the $PL+WN$ stochastic noise model. Correlation between the distribution of the residuals and the Normal ($Corr.$ $Normal$) and  the Levy $\alpha$-stable distribution  ($Corr.$ $Levy$)  }
\label{Table3}
\tiny
\begin{tabular}{|l|l|l|l|l|}
\hline
{\it DRAO (PL+WN)} & {\it(err. in mm) ARMA}  & {\it(err. in mm) FARIMA}  & {\it Corr. Normal } & {\it Corr. Levy } \\
{\it East } & 1.07 $\pm$ 0.01 & 1.10 $\pm$ 0.07 & 0.92 $\pm$ 0.05  & 0.94 $\pm$ 0.05\\
 {\it North} & 1.02 $\pm$ 0.02 & 1.01 $\pm$ 0.01 & 0.93 $\pm$ 0.07 & 0.94 $\pm$ 0.06\\
 {\it Up} & 2.32 $\pm$ 0.21 & 2.15 $\pm$ 0.30 & 0.94 $\pm$ 0.04 & 0.94 $\pm$ 0.05\\
\hline
\hline
 {\it ASCO (PL+WN)} &   &   &  &  \\
 {\it East } & 0.77 $\pm$ 0.01 & 0.77 $\pm$ 0.06 & 0.95 $\pm$ 0.03  & 0.96 $\pm$ 0.05\\
 {\it North} & 0.84 $\pm$ 0.03 & 0.73 $\pm$ 0.03 & 0.97 $\pm$ 0.02 & 0.96 $\pm$ 0.03\\
 {\it Up} & 2.71 $\pm$ 0.12 & 2.34 $\pm$ 0.17 & 0.93 $\pm$ 0.03 & 0.94 $\pm$ 0.01\\
 \hline
 \hline
 {\it ALBH (PL+WN)} &   &  &  &  \\
 {\it East } & 0.97 $\pm$ 0.06 & 0.87 $\pm$ 0.06 & 0.94 $\pm$ 0.01  & 0.94 $\pm$ 0.01\\
 {\it North} & 1.54 $\pm$ 0.03 & 1.06 $\pm$ 0.14 & 0.90 $\pm$ 0.02 & 0.91 $\pm$ 0.04\\
 {\it Up} & 4.36 $\pm$ 0.17 & 4.08 $\pm$ 0.25 & 0.92 $\pm$ 0.05 & 0.94 $\pm$ 0.01\\
 \hline
\end{tabular}
\end{table}

Now Table \ref{Table3} displays the statistics on the error when fitting the ARMA and FARIMA models to the residuals estimated with the PL+WN stochastic noise model. Note that Table \ref{Table3b} displays in the Appendices the results when using the $FN+WN$ stochastic noise model. The FARIMA and ARMA models perform closely for the whole three stations. The large value  for the Up coordinate is due to the amplitude of the noise much larger for this coordinate than for the East and North components \citep{Montillet2012}. In terms of correlating the distribution of the residuals with the Normal and the Levy $\alpha$-stable distribution, the correlation value is relatively the same for all stations which indicates that the distribution of the residuals are Gaussian  with the absence of large tails. Those results further support the selection of the fractional Levy process as the third r.v. However, the study of real time series also underlines the difficulty to characterize statistically this third r.v.
 
\subsection{Discussion on the Limits of Modeling with Levy Processes}\label{Section3.4}
As discussed in the previous sections, the stable Levy process is characterized by a very large (or infinite) variance. In \cite{MontilletYu}, it was assumed that the infinite variance of the residual time series comes from large tails of the distribution (also called heavy tails -\citep{Wooldridge}), generated by a large amplitude of coloured noise, outliers and other remaining geophysical signals. The same study implied that the values of the noise variance should be bounded,
excluding extreme values. This is an important assumption to decide whether
or not (symmetric) $\alpha$-stable distributions can be used to model any geodetic time series. Here, we are investigating how the variance due to residual tectonic rate or seasonal signal evolves with the length of the residual time series (i.e. $L$ epochs).

To recall Section \ref{sec:2.1} and the assumption on the noise properties, let us estimate the mean  and variance of the residual time series. Here, we call the residual time series after the first iteration $\mathbf{s}_1 = [ s_1(t_1), ... , s_1(t_L)]$ $=\Delta^1 \mathbf{s}$ as defined in the previous section. The mean $< s_1(L)>$ and variance $\sigma^2 (L)$ are computed over $L$ epochs (i.e. considering the $L$-th epoch defined as $t_L$ $= Ldt$, with the sampling time $dt$ equal $1$ for simplification and without taking into account any missing epoch in order to have a continuous time series). Based on \cite{Papoulis}, one can estimate the mean over $L$ epochs $< s_1(L)>$ in the cases of remaining linear trend, such as:
\begin{eqnarray}
\label{eqa1}
s_1(t_i) & = & a_r t_i + b_r + n(t_i) \nonumber \\
<s_1(L)> & = & \frac{1}{L} \sum_{i=1}^L (a_r t_i +b_r +n(t_i)) \nonumber \\
<s_1(L)> & = & b_r + a_r \frac{(L+1)}{2} +\mu_C \nonumber \\
<s_1(L)> & \simeq &  a_r \frac{L}{2}  +\mu_C 
\end{eqnarray}
where $a_r$ and $b_r$ are the amplitude and the intersect of the residual trend (i.e. remaining tectonic rate).  Note that the subscript $r$ designates{ \it residual} in the remaining section. $\simeq$ is the approximation for $L >>1$. For a time series with $L$ epochs, the variance $\sigma^2 (L)$ is:
\begin{eqnarray}\label{eqa2}
\sigma^2 (L) &=& \frac{1}{L} \sum_{i=1}^L (s_1(t_i) - < s_1(L)>)^2 \nonumber \\
\sigma^2 (L) &=& a_r^2 \frac{(L + 1)(2L + 1)}{6} - a_r^2 \frac{(L + 1)^2}{4} + b_r^2 + \frac{2a_r}{L} Cross+ \sigma_n^2(L) -\mu_C(\mu_C + a_r (L+1)) \nonumber\\
\sigma^2 (L) &\simeq & \frac{a_r^2 L^2}{12} + \sigma_n^2(L)+ b_r^2 -\mu_C a_r L
\end{eqnarray}
Note that $Cross$ is the cross term between $a_r t_i$ and the noise term $n(t_i)$.
Now, if we assume that the remaining seasonal signal $S_r(t)$ is a pseudo periodic function at frequencies similar to the seasonal signal in Equation \eqref{firsteq}, hence taking the form $S_r(t) = \sum_{j=1}^N c_{r,j}  \cos{(d_j t)} + e_{r,j} \sin{(d_j t)}$. Thus, we can do the same estimation as above in the case of a remaining pseudo periodic component in the residual time series, such as:
\begin{eqnarray}\label{eqa3}
s_1(t_i) &=& S_r(t_i) + n(t_i) \nonumber \\
<s_1(L)> & = & \frac{1}{L} \sum_{i=1}^L (S_r(t_i) +n(t_i)) \nonumber \\
<s_1(L)> & \simeq & \delta + \mu_C 
\end{eqnarray}
where $\delta$ is the average of the remaining seasonal signal. It is assumed to be independent of $L$ and bounded such as a periodic function. The variance is equal to:
\begin{eqnarray}\label{eqa4}
\sigma^2 (L) &=& \frac{1}{L} \sum_{i=1}^L \sum_{j=1}^N c_{r,j}^2  \cos{(d_j t)}^2 + e_{r,j}^2 \sin{(d_j t)}^2 + \sigma_n^2(L) \nonumber \\
 & &+ \frac{2}{L} Cross - {<s_1(L)>}^2 \nonumber \\
 \sigma^2 (L) & \simeq & \sigma_n^2(L) + \sum_{j=1}^N c_{r,j}^2   + e_{r,j}^2  - (\delta + \mu_C )^2
\end{eqnarray}
with $Cross$ is the cross term between $S_r(t)$ and  $n(t)$.  In the Eq. \eqref{eqa1}  to \eqref{eqa4}, the deterministic signals and the noise are assumed completely uncorrelated, which is valid only with white Gaussian noise (i.e. Wiener process) in signal processing \citep{Papoulis}. As previously discussed in Section \ref{sec:2.1}, coloured noise can generate long- memory processes, hence producing non-zero covariance with residual signals. Due to the varying amplitude of the coloured noise in geodetic time series with mixed spectra, the uncorrelated assumption is currently debated within the community \citep{Herring2016,He}. Therefore, recent works have introduced a random component together with a deterministic signal: nonlinear rate \citep{Wang2016,Dimitrieva}, non-deterministic seasonal signal \citep{Davies,Chen,Klos}. Thus, strictly speaking, $\sigma^2$ should be seen as an upper bound.

 The closed-form solution of the variance $\sigma^2 (L)$ shows that the variance is unbounded in the case of a residual linear trend. To recall the discussion in Section \ref{Section3.1}, if this residual trend originates from various sources not well-described in the functional and stochastic model (i.e. undetected jumps, small amplitude random-walk component) of the geodetic time series, the amplitude of this trend should be rather small ($a < 1$ mm/yr) considering the length of GNSS time series available until now ($L < 30$ years). Unless this nonlinear residual trend has a large amplitude, a correction of the functional model must be done a posteriori due to possible anxiety between the models and the observations. The same remarks can be applied to the variance of the remaining seasonal signal where a large amplitude would imply a misfit with the functional model. Thus, we expect rather small amplitude of the coefficients $c_{r,j}$ and $e_{r,j}$ $\sim$  $0.1$ to $\sim 0.001$ mm. Also, in the Appendix $B$, we have develop a similar formula to take into account undetected offsets, where we show that the variance is also bounded. In this case, a large value would mean that one or several large offsets have not been included in the functional model.

%
%
%
\section{Conclusions}
We have investigated the statistical assumptions behind using the fBm and
the family of $\alpha$-stable distributions in order to model the stochastic processes within the residual GNSS time series. 
We model the residual time series as a sum of three r.v. The first two are defined from the stochastic model and assumptions on the noise properties of the geodetic time series. The third r.v. is assumed to belong to the Levy processes. We then distinguish three cases. In the case of a residual time series containing only short-term processes, the r.v. is a Gaussian Levy process. In the presence of long-term correlations and exhibiting self-similarity property, fractional Levy processes can be seen as an alternative model of using the fBm. Due to the linear relationship between the Hurst parameter and the fractional parameter of the FARIMA, it is likely that the FARIMA can fit the residual time series under specific conditions (i.e. amplitude of the coloured noise). The third case is the stable Levy process, with the presence of long-term correlation processes, high amplitude aggregation processes or random-walk.
\\ \indent In order to check our model, we have simulated mixed spectra time series with various levels of coloured noise. We have then developed a $N$ steps methodology based on varying the length of the time series (limited to $1$ year) to study the variations of the stochastic and functional models and to model the distribution of the residuals. The results emphasize the difficulty to separate the fractional Levy process and the stable Levy process mainly due to the absorption of the variations of stochastic processes by the functional model, unless the distribution of the residuals exhibits heavy-tails.  Another difficulty is the dependence of the results with the stochastic noise model. The use of real GNSS time series supports the results based on simulated ones.
\\ \indent However, the discussion on the limits of modeling the stochastic properties of the residuals with the stable Levy process underlines that the infinite variance property can only be satisfied in the case of heavy-tailed distributions.
This condition is generally satisfied if there is a large amplitude random-walk (e.g.,  temporal aggregation in financial time series) or an important misfit between the models (i.e. functional and stochastic) and the observations, which means that there is anxiety in the choice of the functional model (e.g., unmodeled large jumps, large outliers). 
Finally, with longer and longer time series, one may be able to statistically characterize more precisely the third r.v.

\begin{acknowledgements}
We would like to thank Dr. Machiel S. Bos from the SEGAL for multiple discussions on the stochastic properties of the GNSS time series. Dr. Xiaoxing He was sponsored by the Doctoral Fund of Ministry of Education of China (2018M632909) and the National Natural Science Foundation of China (41574031).
\end{acknowledgements}
\vspace{3em}
\appendix{Appendix A}
\section*{fBm and fLsm: integral representation}
The fractional Brownian motion (fBm) $\{B_H (t)\}_{t\geq0}$ has the integral representation:
\begin{equation}
B_H(t)= \int_{-\infty}^{\infty}  \big ( (t-u)_{+}^{H - \frac{1}{2}} - (-u)_{+}^{H - \frac{1}{2}}  \big ) dB(u)
\end{equation}
where $x_+ = max(x,0)$ and $B(u)$ is a Brownian motion (Bm). It is $H$-self-similar with stationary increments and it is the only Gaussian process with such properties for $0 < H < 1$ \citep{Samorodnitsky}.

\vspace{0.5em}
  From  \cite{Weron2005}, the fractional Levy stable motion (fLsm) can be defined with the process $\{Z_\alpha^H (t)\}$ (with $t$ in $\mathbb{R}$) by the following integral representation:

\begin{equation}
Z_\alpha^H= \int_{-\infty}^\infty \big ( (t-u)_{+}^{H - \frac{1}{\alpha}} -(-u)_{+}^{H - \frac{1}{\alpha}} \big ) dZ_\alpha(u)
\end{equation} 
where $Z_\alpha(u)$  is a symmetric Levy $\alpha$ -stable motion (Lsm). The integral is well defined for $0 < H < 1$ and $0 < \alpha \leq 2$ as a weighted average of the Levy stable motion $Z_\alpha(u)$. The process $\{Z_\alpha^H (t)\}$ is H-self-similar and has stationary increments. Comparing the definition of fBm and fLsm, we can observe that fLsm is similar to fBm for the case $\alpha=2$.
%
\vspace{5em}
\appendix{Appendix B}
\section*{Estimation of the Variance in the Presence of Offsets}
We model here the offsets in the time series as Heaviside step functions according to \cite{He}. Following Section \ref{Section3.4}, the residual time series in presence of remaining offsets can be written such as
\begin{equation}
s_1(t_i)= \sum_{k=1}^{ng} g_k \mathcal{H}(t_i-T_k) + n(t_i)
\end{equation}
Where $\mathcal{H}$ is the Heaviside step function. One can estimate the mean over $L$ epochs:
\begin{eqnarray}
<s_1(L)> &=& \frac{1}{L} \sum_{i=1}^L(\sum_{k=1}^{ng} g_k \mathcal{H}(t_i-T_k) ) +\mu_C \nonumber \\
<s_1(L)> &=& \frac{1}{L}\sum_{k=1}^{ng} g_k \mathcal{H}(t_L-T_k) +\mu_C
\end{eqnarray}
The variance is equal to
\begin{eqnarray}
\sigma^2 (L) &=&  \frac{1}{L}\sum_{i=1}^L(\sum_{k=1}^{ng} g_k \mathcal{H}(t_i-T_k) + n(t_i) - {<s_1(L)>})^2 \nonumber \\
\sigma^2 (L) &\simeq & \sigma_n^2(L) + \frac{1}{L}(\sum_{k=1}^{ng} g_k \mathcal{H}(t_L-T_k))^2 - {<s_1(L)>}^2
\end{eqnarray}
In the presence of small  (undetectable) offsets ( $g_k < 1$ mm),  we can further assume that $<s_1(L)> \sim \mu_C$ and $\sigma^2 (L)  \sim \sigma_n^2(L) -\mu_C^2$.  For multiple large uncorrected offsets (i.e. noticeable above the noise floor),  the variance can be large, but the distribution of the residual time series should look like various Gaussian distributions overlapping each other corresponding to the segments of the time series defined by those noticeable offsets. This case is not taken into account in our assumptions summarized in Table \ref{Table1}, because it supposes that there is a large anxiety about the chosen functional model - obviously missing some large offsets.  
\newpage
\appendix{Appendix C}
\section*{Additional Results}


\begin{figure}[hp!]
 \caption{Percentage of variations of the estimated parameters included in the stochastic and functional models when varying the length of the daily position GNSS time series corresponding to the stations $DRAO$, $ASCO$ and $ALBH$. The statistics are estimated over the East, North and Up Corrdinates \label{Figure1XX}}
 \hspace{-10em}
 \includegraphics[width=1.7\textwidth]{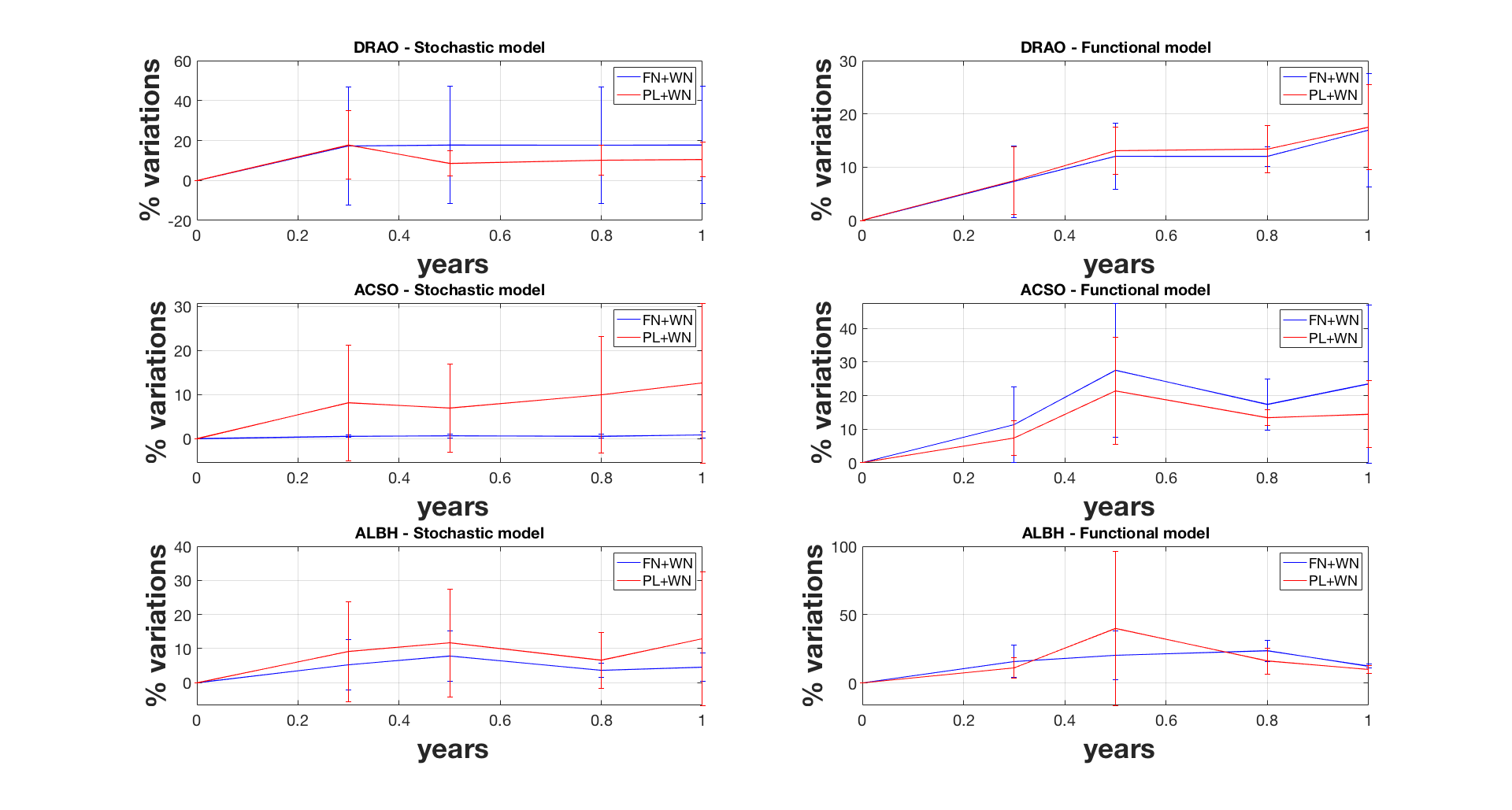}
\end{figure} 

\begin{table}[!htbp]
 \centering 
\caption{Statistics on the Error when fitting the ARMA and FARIMA model to the residual time series for each coordinate of the stations $ALBH$, $DRAO$ and $ASCO$ based on the $FN+WN$ stochastic noise model. Correlation between the distribution of the residuals and the Normal ($Corr.$ $Normal$) and  the Levy $\alpha$-stable distribution  ($Corr.$ $Levy$)  }
\label{Table3b}
\tiny
\begin{tabular}{|l|l|l|l|l|}
\hline
{\it DRAO ($FN+WN$)} & {\it(err. in mm) ARMA}  & {\it(err. in mm) FARIMA}  & {\it Corr. Normal } & {\it Corr. Levy } \\
{\it East } & 1.07 $\pm$ 0.01 & 1.00 $\pm$ 0.02 & 0.92 $\pm$ 0.03  & 0.94 $\pm$ 0.05\\
 {\it North} & 1.02 $\pm$ 0.02 & 1.32 $\pm$ 0.07 & 0.92 $\pm$ 0.05 & 0.94 $\pm$ 0.04\\
 {\it Up} & 2.33 $\pm$ 0.18 & 2.20 $\pm$ 0.32 & 0.94 $\pm$ 0.08 & 0.94 $\pm$ 0.05\\
\hline
\hline
 {\it ASCO ($FN+WN$)} &   &   &  &  \\
 {\it East } & 0.77 $\pm$ 0.01 & 0.75 $\pm$ 0.07 & 0.95 $\pm$ 0.02  & 0.96 $\pm$ 0.01\\
 {\it North} & 0.85 $\pm$ 0.03 & 0.74 $\pm$ 0.05 & 0.94 $\pm$ 0.01 & 0.96 $\pm$ 0.01\\
 {\it Up} & 2.18 $\pm$ 0.14 & 2.51 $\pm$ 0.21 & 0.93 $\pm$ 0.03 & 0.94 $\pm$ 0.03\\
 \hline
 \hline
 {\it ALBH ($FN+WN$)} &   &  &  &  \\
 {\it East } & 0.97 $\pm$ 0.04 & 0.86 $\pm$ 0.06 & 0.93 $\pm$ 0.01  & 0.94 $\pm$ 0.01\\
 {\it North} & 1.52 $\pm$ 0.08 & 1.08 $\pm$ 0.10 & 0.91 $\pm$ 0.02 & 0.91 $\pm$ 0.04\\
 {\it Up} & 3.83 $\pm$ 0.21 & 3.32 $\pm$ 0.15 & 0.93 $\pm$ 0.03 & 0.94 $\pm$ 0.01\\
 \hline
\end{tabular}
\end{table}


\end{document}